\input psfig.sty

\def\ptitle{Coulomb plus power-law potentials in quantum mechanics}  
\nopagenumbers 
\magnification=\magstep1 
\hsize 6.0 true in 
\hoffset 0.25 true in 
\emergencystretch=0.6 in 
\vfuzz 0.4 in 
\hfuzz 0.4 in 
\vglue 0.1true in 
\mathsurround=2pt 
\def\nl{\noindent} 
\def\nll{\hfil\break\noindent} 
\def\np{\hfil\vfil\break} 
\def\ppl#1{{\noindent\leftskip 9 cm #1\vskip 0 pt}} 
\def\title#1{\bigskip\noindent\bf #1 ~ \trr\smallskip} 
 

\def\frac#1#2{{#1\over#2}}

\font\trr=cmr10    
\font\bf=cmbx10    
\font\sl=cmsl10    
\font\it=cmti10    
\font\trbig=cmbx10 scaled 1500 
\font\tiny=cmr8    
\def\ng{>\kern -9pt|\kern 9pt} 
\def\hi#1#2{$#1$\kern -2pt-#2} 
\def\hy#1#2{#1-\kern -2pt$#2$} 

\def\sgn{{\rm sgn}}

\def\nl{\noindent}    
\def\nll{\hfil\break} 

\output={\shipout\vbox{\makeheadline\ifnum\the\pageno>1 {\hrule} \fi 
{\pagebody}\makefootline}\advancepageno} 
 
\headline{\noindent {\ifnum\the\pageno>1 
{\tiny \ptitle\hfil page~\the\pageno}\fi}} 
\footline{} 
\newcount\zz \zz=0 
\newcount\q 
\newcount\qq \qq=0 
 
\def\pref #1#2#3#4#5{\frenchspacing \global \advance \q by 1 
\edef#1{\the\q}{\ifnum \zz=1 { %
\item{[\the\q]}{#2} {\bf #3},{ #4.}{~#5}\medskip} \fi}} 
 
\def\bref #1#2#3#4#5{\frenchspacing \global \advance \q by 1 
\edef#1{\the\q}{\ifnum \zz=1 { %
\item{[\the\q]}{#2}, {\it #3} {(#4).}{~#5}\medskip} \fi}} 
 
\def\gref #1#2{\frenchspacing \global \advance \q by 1 
\edef#1{\the\q}{\ifnum \zz=1 { %
\item{[\the\q]}{#2}\medskip} \fi}}

\def\sref #1{~[#1]} 
 
\def\references#1{\zz=#1 
\parskip=2pt plus 1pt 
{\ifnum \zz=1 {\noindent \bf References \medskip} \fi} \q=\qq

%


\pref{\varma} {S. N Biswas, K. Datt, R. P. Saxena, P. K. Strivastava, and V. S. Varma, J. Math. Phys., No. 9}{14}{1190 (1972)}{}
\pref{\castro }{Francisco M., Ferndez and Eduardo A. Castro, Am. J. Phys., No. 10}{50}{921 (1982)}{}
\bref{\barn}{J. F. Barnes, H. J. Brascamp, and E. H. Lieb}{In: Studies in Mathematical Physics: Essays in Honor of Valentine Bargmann (Edited by E. H. Lieb, B. Simon, and A. S. Wightman)}{Princeton University Press, Princeton, 1976}{p 83}
\pref{\hio} {F. T. Hioe, Don MacMillen, and E. W. Montroll, J. of Math. Phys., No 7}{17}{(1976)}{}
\pref{\freu}{H. Freutus, Lett. Nuovo Cim.}{22}{26 (1976)}{}
\pref{\trus}{H. Turschner, J.Phys. A, No. 4}{12}{451 (1978)}{}
\pref{\nieto }{ M. M. Nieto and L. M. Simons, Am. J. Phys.}{47}{634 (1979)}{}
\pref{\hill}{B. J. B. Crowley and T. F. Hill, J. Phys. A}{12}{223 (1979)}{}
\pref{\mark}{Mark S. Ashbaugh and John D. Morgan III, J. Phys. A}{14}{809 (1981)}{}
\pref{\reno}{R. E. Carndall and Mary Hall Reno, J. Math. Phys. }{23}{ 64 (1982) }{}
\pref{\rhallcp}{ R.L Hall, Phys. Rev. D}{30 }{433 (1984)  }{ }
\pref{\hallpow}{R. L. Hall, Phys. Rev. A}{39}{5500 (1989)}{}
\pref{\hallpn}{R. L. Hall and N. Saad, J. Math. Phys.}{38}{4909 (1997)}{}
\pref{\god}{S. Godfrey and J. Napolitano, Rev. Mod. Phys.}{71}{1411 (1999)}{}
\pref{\alhai}{A. D. Alhaidari, Int. J. Mod. Phys. A}{17}{4551 (2002)}{}
\pref{\hallenva}{R. L. Hall, Phys. Rev. D} {22}{2062-2072 (1980)}{}
\pref{\hallenv}{R. L. Hall, J. Math. Phys.}{34}{2779 (1993)}{}
\pref{\rhallsum}{R.L Hall, J. Math. Phys.}{33}{1710 (1992) }{}
\pref{\comp}{R. L. Hall and Q. D. Katatbeh, J. Phys. A}{35}{8727 (2002)}{}
\pref{\cif}{H. Ciftci, E. Ate\c{s}er and H. Koru, J. Phys. A}{36}{3821 (2003)}{}



} 
 
\references{0} 
\topskip 2pt
\trr 
 \ppl{CUQM-96}\ppl{math-ph/0305018}\medskip 
\vskip 0.6 true in 
\centerline{\trbig \ptitle}
\vskip 0.4true in
\baselineskip 12 true pt 
\centerline{\bf Hakan Ciftci$^*$, Richard L. Hall$^{\dag}$ and Qutaibeh D. Katatbeh$^{\dag}$}\medskip
\centerline{\sl $^{\dag}$ Department of Mathematics and Statistics,}
\centerline{\sl Concordia University,}
\centerline{\sl 1455 de Maisonneuve Boulevard West,}
\centerline{\sl Montr\'eal, Qu\'ebec, Canada H3G 1M8.}
\vskip 0.2 true in
\centerline{\sl $^*$Gazi Universitesi, Fen-Edebiyat Fak\"{u}ltesi}
\centerline{\sl Fizik B\"{o}l\"{u}m\"{u}. 06500 Teknikokullar}
\centerline{ Ankara, Turkey} 
\vskip 0.2 true in
\centerline{email:\sl~~rhall@mathstat.concordia.ca}
\bigskip\bigskip
\baselineskip = 18true pt 
 
\centerline{\bf Abstract}\medskip 

We study the discrete spectrum of the Hamiltonian $H=-\Delta+V(r)$ for the Coulomb plus power-law potential $V(r)=-1/r+\beta\ \sgn (q)r^q,$ where $\beta>0,\ q>-2\ {\rm and }\ q\ne 0$. We show by envelope theory that the discrete eigenvalues $E_{n\ell}$  of $H$ may be approximated by the semiclassical expression $E_{n\ell}(q)\approx\min_{r>0}\{1/r^2-1/(\mu r)+ \sgn(q)\beta(\nu r)^q\}.$ Values of $\mu$ and $\nu$ are prescribed which yield upper and lower bounds. Accurate upper bounds are also obtained by use of a trial function of the form, $\psi(r)= r^{\ell+1}e^{-(xr)^{q}}$. We give detailed results for $V(r)=-1/r+\beta r^q,\ q=0.5,1,2$ for $n=1,\ \ell=0,1,2,$ along with comparison eigenvalues found by direct numerical methods.   

\nll PACS: 03.65.Ge 
\np 
\topskip 20pt
\title{1.~~Introduction} 

In this paper we derive upper and lower bound formulas for the spectrum of a single particle in three dimensions that obeys non-relativistic quantum mechanics and has Hamiltonian 
$$H=-\omega\Delta -A/r+B\sgn(q)r^q,\quad \omega,\ A,\  B >0,\ {\rm and} \ q\ne 0,\ q>-2.\eqno{(1.1)}$$ 
The Coulomb plus power-law potential is of interest in particle physics where it serves as a non-relativistic model for principle part of the quark-quark interaction.  This class of potentials has been well studied and much work has been done to approximate the eigenvalues, with or without the Coulomb term necessitated by QCD\sref{\varma-\alhai}. Our goal in this paper is to provide simple formulas for upper and lower energy bounds for this class of potentials. Firstly, we use the `envelope method'\sref{\hallenva,\hallenv} to obtain upper and lower bound formulas for all the discrete eigenvalues. We also use a Gaussian trial function and the `sum approximation'\sref{\rhallsum,\comp} to improve the bounds for the bottom of each angular-momentum subspace. The energy bounds so far discussed may all be expressed in terms of the following semiclassical energy formula:
$${\cal E}\approx \min _{r>0}\left \{ \omega{1\over r^2}-{A\over \mu r}+ B\sgn(q) (\nu r)^q \right\}\quad \eqno{(1.2)} $$
\nl for suitable choices of the parameters $\mu>0$ and $\nu>0.$  We also apply a variational method used earlier\sref{\cif} which is based on the exact Coulomb wave function and yields accurate upper bounds for the bottom of each angular momentum subspace. We compare all these results with `exact' eigenvalues computed by direct numerical integration. 
\par For the class of potentials studied some exactly solvable cases exist for suitable values of the couplings $\omega,\ A,\ B,$ and the power $q.$ For example, for the well-known hydrogenic atom and the harmonic oscillator  potentials we have explicitly for $n = 1,2,3, ...$
$$q = -1\quad\Rightarrow\quad E_{n\ell}=-{A^2\over 4\omega(n+\ell)^{2}}\eqno{(1.3)}$$
\nl and 
$$q = 2\quad\Rightarrow\quad E_{n\ell}=(\omega B)^{1\over 2}(4n+2\ell-1).\eqno{(1.4)}$$
\nl For $\ell = 0,$ exact solutions are also available for the linear potential $q = 1.$ We can simplify the coupling problem in general by the use of scaling arguments. If, for each fixed $q,$ we denote the eigenvalues of $H=-\omega \Delta -A/r+Br^{q}$ by ${\cal E}(\omega,A,B)$, and consider a scale change of the form $s=r/{\sigma}$, and choose the scale ${\sigma}={\omega}/A,$ then it is straightforward to show that, 
$$ {\cal E}(\omega,A,B)=\left({A^2\over \omega}\right) {\cal E}(1,1,\beta),\quad \beta = \left({B\over{\omega}}\right)\left({{\omega}\over A}\right)^{q+2}.\eqno{(1.5)} $$
\noindent Hence, the full problem is now reduced to the simpler one-parameter problem
$$ H=-\Delta -1/r+\beta\ \sgn(q)r^q, \quad{E}={E}(\beta) = {\cal E}(1,1,\beta),\ \beta> 0. \eqno{(1.6)}$$ 

\title{2.~~Energy bounds by the envelope method and the sum approximation } 

The comparison theorem tells us that an ordering between potentials implies an  ordering between the corresponding eigenvalues. The `envelope method'\sref{\hallenva, \hallenv} is based on this theorem and establishes upper and lower bound formulas for a wide class of attractive spherically-symmetric potentials. We need a solvable model $-\Delta + h(r)$  which provides an `envelope basis' for the study of the problems of the form $-\Delta + g(h(r)),$ where the transformation function $g$ is monotone increasing and of definite convexity: when $g$ is convex, we obtain lower bounds; when $g$ is concave, the theory yields upper bounds. The natural basis in this contex is a  single power-law potential. The spectrum of a Hamiltonian of the form 
$$H=-\Delta+ \sgn(q)r^q,\ {\rm where}\ q>-2\ {\rm and } \ q \ \neq 0 \eqno{(2.1)}$$
may be represented {\it exactly}  by the following semiclassical expression\sref{\hallpow, \hallenv},
$$E_{n\ell}=\min_{r>0}\left \{ {1\over r^2} +\sgn(q) (P_{n\ell}(q)r)^q \right\}\quad \eqno{(2.2a)}$$ 
$$= \sgn(q)\left(1+{q\over 2}\right)\left({{2P_{n\ell}(q)^{2}}\over {|q|}}\right)^{q\over {2+q}}  .\eqno{(2.2b)}$$ 
The function $P=P_{n\ell}(q)$ is known as the \hi{P}{representation}, for the Schr\"odinger spectra generated by the power-law potentials. It is convenient to use the $P$ function to study and analyse the spectra of these problems mainly because it is known\sref{\hallpow} that $P_{n\ell}(q)$ is monotone in $q$ and it is also smoother than $E_{n\ell}$ as a function of $q;$ the case $q = 0$ corresponds exactly to the $\log$ potential. From (1.3) and (1.4) we find, in particular, that:
$$P_{n\ell}(-1)=n+\ell \eqno{(2.3)}$$
\nl and 
$$P_{n\ell}(2)=2n+\ell+1/2.\quad \eqno{(2.4)}$$
\nl In Table~1 we exhibit some numerical values for $P_{n\ell}({1\over 2})$ and $P_{n\ell}(1).$ We have found the exact eigenvalues for the linear potential in terms of the zeros of the Airy function, but those for $q = {1\over 2}$ have to be computed numerically: this use of some isolated numerical input is justified since, for each $\{n,\ell\}$ pair, the resulting approximation formulas include all the potential parameters but depend only on a single `numerical input'.
 Envelope theory\sref{\hallpn, \rhallsum}
 shows that the eigenvalues of the Coulomb plus power-law potential may be approximated by the following semiclassical expression, 
$${\cal E}\approx \min _{r>0}\left \{ {1\over r^2}-{1\over \mu r}+\beta\ \sgn(q) (\nu r)^q \right\}, \ {\rm where}\  \mu,\ \nu>0.\quad \eqno{(2.5)} $$
\nl Since $V(r) = g(h(r))$ is at once a convex function of $h(r) = -1/r$ and a concave function of $h(r) = \sgn(q)r^q,$ the spectral representation $P_{n\ell}(q)$ allows us to specify upper and lower bounds formulas as follows. If $\mu=\nu=P_{n\ell }(-1) $, then ${\cal E}$ is a lower bound for $E_{n\ell}$, and if $\mu=\nu=P_{n\ell}(q),$ then ${\cal E}$ is an upper bound.  We may improve the lower bound for the bottom of each angular momentum subspace by using the sum approximation\sref{\rhallsum,\comp}, which is equivalent to the choice $\mu=P_{1\ell}(-1)=(\ell+1)$ and $\nu=P_{1\ell}(q).$ For the bottom of the spectrum we can also improve the upper bound by using a Gaussian trial function and minimizing over scale: this is equivalent\sref{\hallpn} to using the parameter values
$$\mu = \nu = P^U_{1 0}= \left({3\over 2}\right)^{{1\over2}} \left[2\Gamma((3+q)/2)\over \sqrt{\pi} \right]^{{1\over q}}.\eqno{(2.6)}$$
\nl We note that the {\it same} parameters $\mu$ and $\nu$ which guarantee that (2.5) yields various energy bounds may also be used in the `full' semiclassical formula (1.2),
including all the original Hamiltonian parameters $\{\omega, A, B\}.$ In Section~3 we apply (2.5) to the explicit cases $V(r)=-1/r+\beta\ r^{q}$ for $\ell=0,1,2,$ where $q=1,2,\ {\rm and }\ 0.5$. 

\title{3.~~Variational method} 

The second approach in this paper is to use a trial function explored in previous work\sref{\cif} to obtain accurate upper bounds for the bottom of each angular momentom subspace. We start with Schr\"odinger's equation
$$H\psi(r)= \left(-\Delta-\frac{1}{r}+\beta\ \sgn(q) r^q\right)\psi(r)={E}_{n\ell}(\beta)\psi(r),\quad q\neq 0,\ q>-2.\eqno{(3.1)}$$
This problem is solvable if $\beta=0$, and the corresponding wave function $\psi(r)$ is given by 
$$\psi(r)= r^{\ell+1}e^{-xr}L_{n}^{2\ell+1}(2xr).\eqno{(3.2)}$$
In order to obtain an upper bound for the bottom of each angular momentum subspace ${ E}_{1\ell}$ for fixed power $q$ we choose $\psi(r)$ to be of the following form
$$\psi(r)= r^{\ell+1}e^{-(xr)^{d}}\eqno{(3.5)}$$
and define ${\cal E}$ by ${\cal E}(\beta,x,d)={(\psi,H\psi)\over (\psi,\psi)},$ where $x$ and $d$ are variational parameters. Now, we minimize ${\cal E}$ with respect to $x$ and $d.$ The necessary conditions for a critical point are
$\frac{\partial{\cal E}}{\partial x}=0$ and $\frac{\partial{\cal E}}{\partial d}=0.$
Consequently, using (3.1) and (3.5), we obtain the following upper bound formula for the eigenvalues ${E}_{1\ell}$
$${\cal E}_{1\ell}(\beta,d,x)=a_{1}x^{2}-a_{2}x+a_{3}x^{-q},\eqno{(3.6)}$$
where $a_{1}$, $a_{2}$ and $a_{3}$ are as given below%
$$a_{1}=2^{\frac{2-2d}{d}}\frac{(2\ell+1)(2\ell+d+1)\Gamma(\frac{2\ell+1}{d})}{\Gamma(\frac{2\ell+3}{d}
)}$$
$$a_{2}=2^{\frac{1}{d}}\frac{\Gamma(\frac{2\ell+2}{d})}{\Gamma(\frac{2\ell+3}{d})}$$
$$a_{3}=\sgn(q)\beta2^{\frac{-q}{d}}\frac{\Gamma(\frac{2\ell+q+3}{d})}
{\Gamma(\frac{2\ell+3}{d})}.$$
\nl By using (3.6) we derive the following equation for $x$%
$$x^{q+2}-\frac{a_{2}}{2a_{1}}x^{q+1}-\frac{qa_{3}}{2a_{1}}=0.\eqno{(3.7)}$$
\nl After solving Eq.(3.7) to obtain $x,$  from the numerical solution of 
$\frac{\partial \varepsilon }{\partial d}=0$
 we find $d$ for $n=1$ and $\ell=0,$ and then we use the same $d$ 
value for all $\ell.$   

 \title{4.~~Results and Conclusion}

We have found general semiclassical energy formulas (1.2) and (2.5) for the eigenvalues generated by the Coulomb plus power-law potentials. Specific values for the parameters $\mu$ and $\nu$ are given which guarantee that the formulas yield bounds for all the discrete energies. By using a more finely tuned wavefunction, we have also derived an improved upper bound (3.6) valid for the bottom of each angular momentum subspace. We may rewrite (2.5) in the form of a pair of parametric equations for the curve $\{\beta,\ E(\beta)\}$. For fixed $q > -1$ we obtain:
 
$${\eqalign{ \beta&= {1\over{|q|(\nu r)^{q}}}\left({2\over{r^2}}-{1\over{\mu r}} \right)\cr\cr
E(\beta)&= {{1+2/q}\over {r^2}} -{{1 +1/q}\over {\mu r}} .\cr}}\eqno{(4.1)}$$

\nl By envelope theory, we know that these parametric equations yield a lower bound if $\mu=\nu=P_{n\ell }(-1) = (n+\ell),$ and an upper bound when $\mu=\nu=P_{n\ell}(q).$  For the bottom of each angular momentum subspace the prescription $\mu=P_{1\ell}(-1)=(\ell+1),$ $\nu = P_{1\ell}(q)$ yields an improved lower bound.  An improved upper bound for the bottom of the spectrum is given by using the `Gaussian' \hi{P}{numbers} (2.6). In Figures~1, 2, and 3, we plot the function $E(\beta)$ for $n=1\ ,\ell=0,1,2$ for the Coulomb plus harmonic oscillator $(q = 2)$, Coulomb plus linear $(q = 1),$ and Coulomb plus $r^{0.5}$ potentials, along with the corresponding accurate variational bounds using (3.6) (dashed line), and some comparison numerical values represented as stars.  The advantage of the semiclassical formulas is that that they describe in approximate analytical form how the eigenvalues depend on all the parameters of the problem.

\title{Acknowledgment} 
Partial financial support of this work under Grant No. GP3438 from the Natural 
Sciences and Engineering Research Council of Canada is gratefully acknowledged
by one of us [RLH]. 

\np
\vskip 0.4in
  

\noindent {\bf Table 1}~~The `input' values $P_{n\ell}({1\over 2})$ and $P_{n\ell}(1)$ to be used in the general formula (2.5) for the energies corresponding to the potential $V(r) = -1/r + \beta\  \sgn(q)r^q.$  These \hi{P}{values} yield upper bounds when $q \leq {1\over 2},$ or $q\leq 1,$ respectively.   
\baselineskip=16 true pt 
\def\vr{\vrule height 12 true pt depth 6 true pt}
\def\vra{\vr\hfill} \def\vrb{\hfill &\vra} \def\vrc{\hfill & \vr\cr\hrule}
\def\vrq{\vr\quad} 
$$\vbox{\offinterlineskip
 \hrule
\settabs
\+ \vrq \kern 0.4true in &\vrq \kern 0.4true in &\vrq \kern 0.7true in &\vrq \kern 0.7true in &\vrq \kern 0.7true in &\vr\cr\hrule
\+ \vra $n$ \vrb $\ell$\vrb $P_{n\ell}({1\over 2})$\vrb  $P_{n\ell}(1)$ \vrc
\+ \vra 1\vrb 0\vrb  1.30266\vrb  1.37608\vrc
\+ \vra 2\vrb 0\vrb  2.97387\vrb  3.18131\vrc
\+ \vra 3\vrb 0\vrb 4.65440\vrb  4.99255\vrc
\+ \vra 4\vrb 0\vrb  6.33742\vrb  6.80514\vrc
\+ \vra 5\vrb 0\vrb  8.02149\vrb  8.61823\vrc
\+ \vra 1\vrb 1\vrb 2.29747\vrb  2.37192\vrc
\+ \vra 2\vrb 1\vrb 3.93966\vrb  4.15501\vrc
\+ \vra 3\vrb 1\vrb 5.60154\vrb  5.95300\vrc
\+ \vra 4\vrb 1\vrb 7.27194\vrb  7.75701\vrc
\+ \vra 5\vrb 1\vrb 8.94679\vrb  9.56408\vrc
\+ \vra 1\vrb 2\vrb 3.29535\vrb  3.37018\vrc
\+ \vra 2\vrb 2\vrb 4.92261\vrb  5.14135\vrc
\+ \vra 3\vrb 2\vrb  6.57089\vrb  6.92911\vrc
\+ \vra 4\vrb 2\vrb 8.23022\vrb  8.72515\vrc
\+ \vra 5\vrb 2\vrb 9.89619\vrb  10.52596\vrc
\+ \vra 1\vrb 3\vrb 4.29424\vrb  4.36923\vrc
\+ \vra 2\vrb 3\vrb 5.91240\vrb  6.13298\vrc
\+ \vra 3\vrb 3\vrb 7.55077\vrb  7.91304\vrc
\+ \vra 4\vrb 3\vrb 9.20118\vrb  9.70236\vrc
\+ \vra 5\vrb 3\vrb 10.85929\vrb  11.49748\vrc
\+ \vra 1\vrb 4\vrb  5.29352\vrb  5.36863\vrc
\+ \vra 2\vrb 4\vrb  6.90560\vrb  7.12732\vrc
\+ \vra 3\vrb 4\vrb 8.53658\vrb  8.90148\vrc
\+ \vra 4\vrb 4\vrb 10.17964\vrb  10.68521\vrc
\+ \vra 5\vrb 4\vrb 11.83110\vrb  12.47532\vrc

}$$

\np

\hbox{\vbox{\psfig{figure=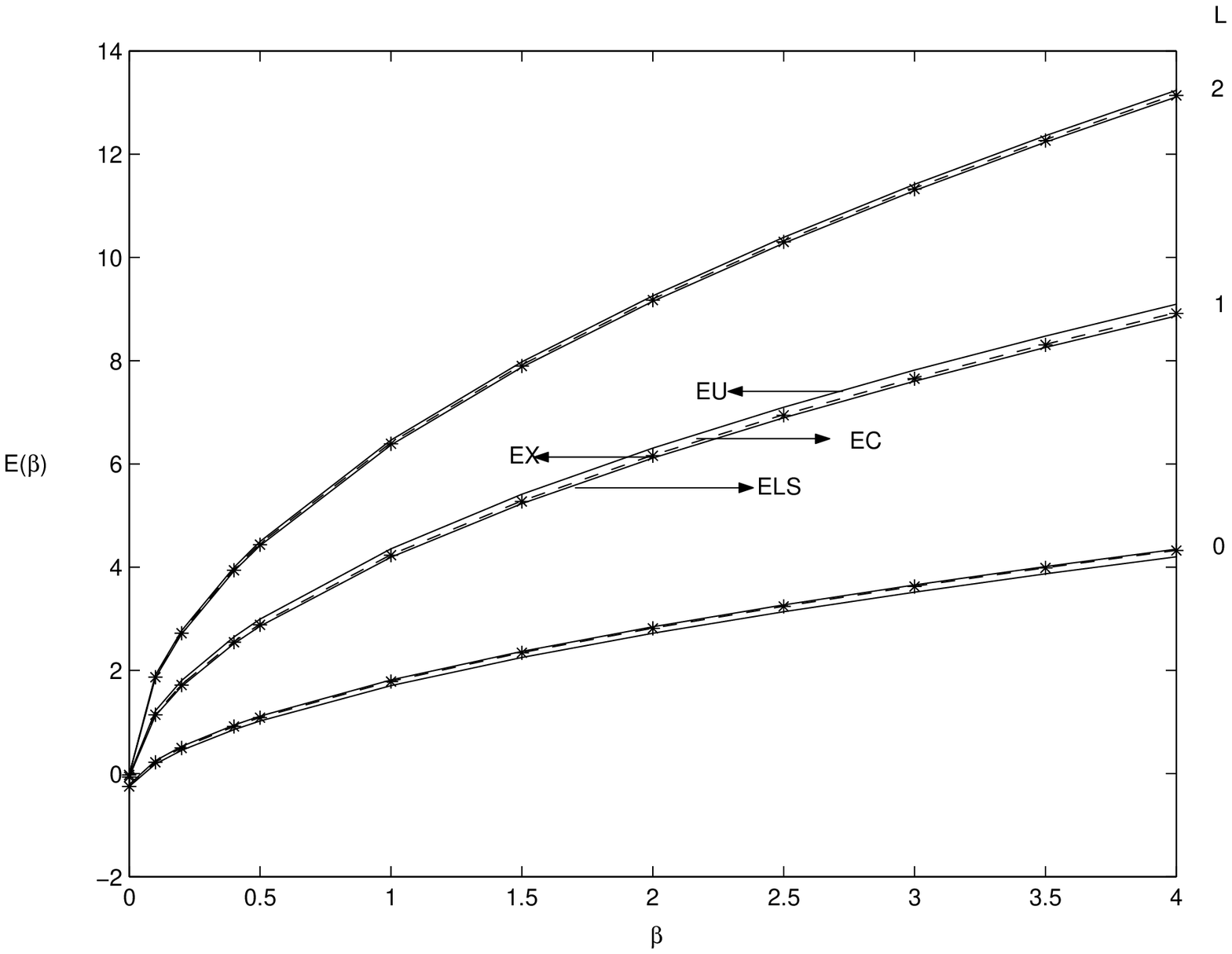,height=4in,width=5in,silent=}}}

\title{Figure 1.}

\nl The eigenvalues ${E}(\beta)$ of the Hamiltonian $H=-\Delta-1/r+\beta r^2$ for dimension $N = 3,$ $n=1,$ and $\ell=0,1,2$. The continuous curves show the upper bound EU given by the envelope formula (2.5) 
with $\nu = \mu = P_{1\ell}(2),$ for $\ell=1,2$ and the lower bound ELS by the sum approximation 
given by the same formula but with $\nu = P_{1\ell}(2)$ and $\mu = P_{1\ell}(-1).$ The upper bound for $\ell=0$ is calculated using $\nu = P_{1\ell}^{U}(2)$ and $\mu = P_{1\ell}^{U}(-1)$ in formula (2.5). The dashed curve $EC$ represents the upper bound by formula (3.6). The stars EX represent accurate numerical data.

\np
\hbox{\vbox{\psfig{figure=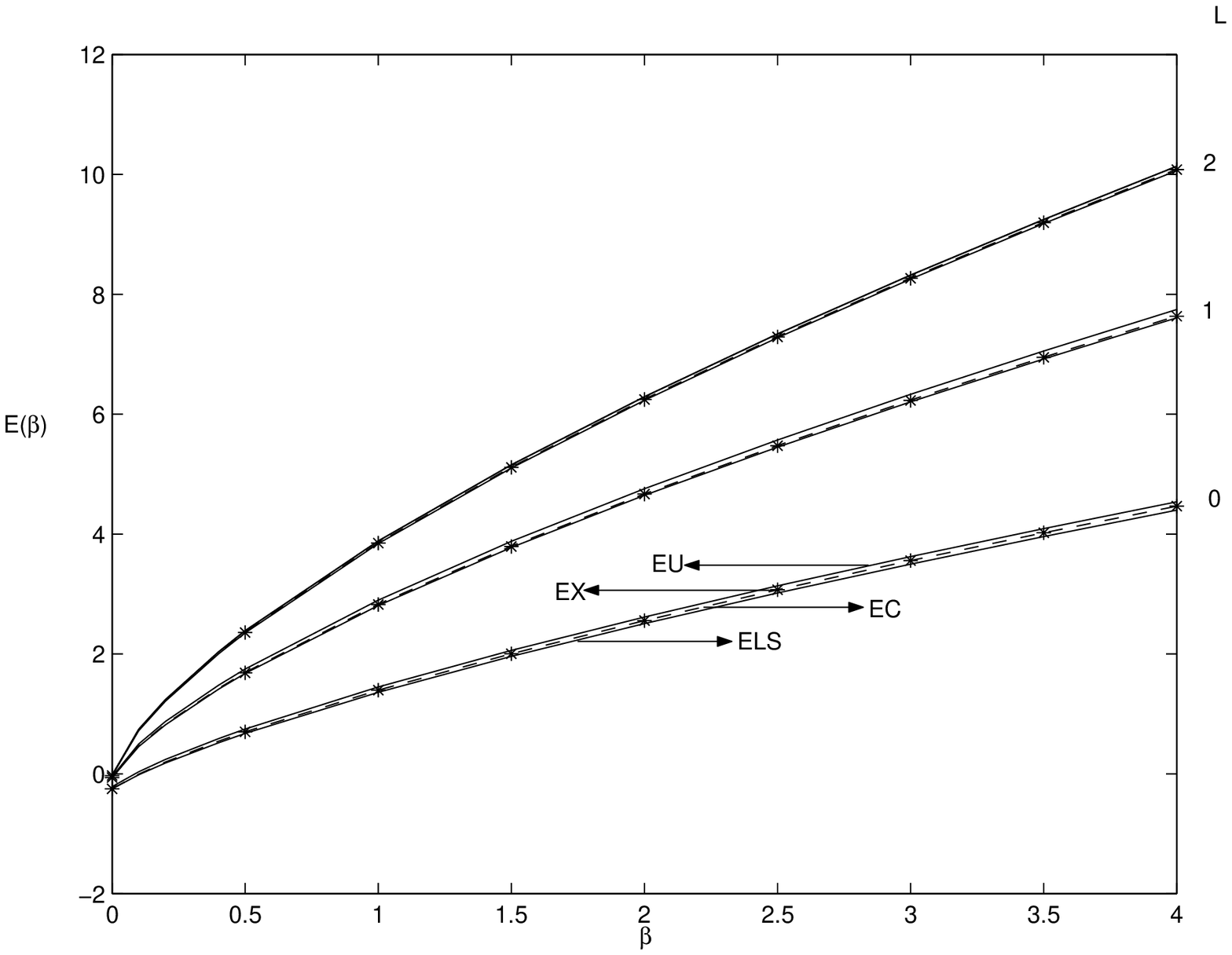,height=4in,width=5in,silent=}}}

\title{Figure 2.}

\nl The eigenvalues ${E}(\beta)$ of the Hamiltonian $H=-\Delta-1/r+\beta r$ for dimension $N = 3,$ $n=1,$ and $\ell=0,1,2$. The continuous curves show the upper bound EU given by the envelope formula (2.5) 
with $\nu = \mu = P_{1\ell}(2),$ and the lower bound ELS by the sum approximation 
given by the same formula but with $\nu = P_{1\ell}(1)$ and $\mu = P_{1\ell}(-1).$ The upper bound for $\ell=0$ is calculated using $\nu = P_{1\ell}^{U}(1)$ and $\mu = P_{1\ell}^{U}(-1)$ in formula (2.5).  The dashed curve $EC$ represents the upper bound by formula (3.6). The stars EX represent accurate numerical data.

\np

\hbox{\vbox{\psfig{figure=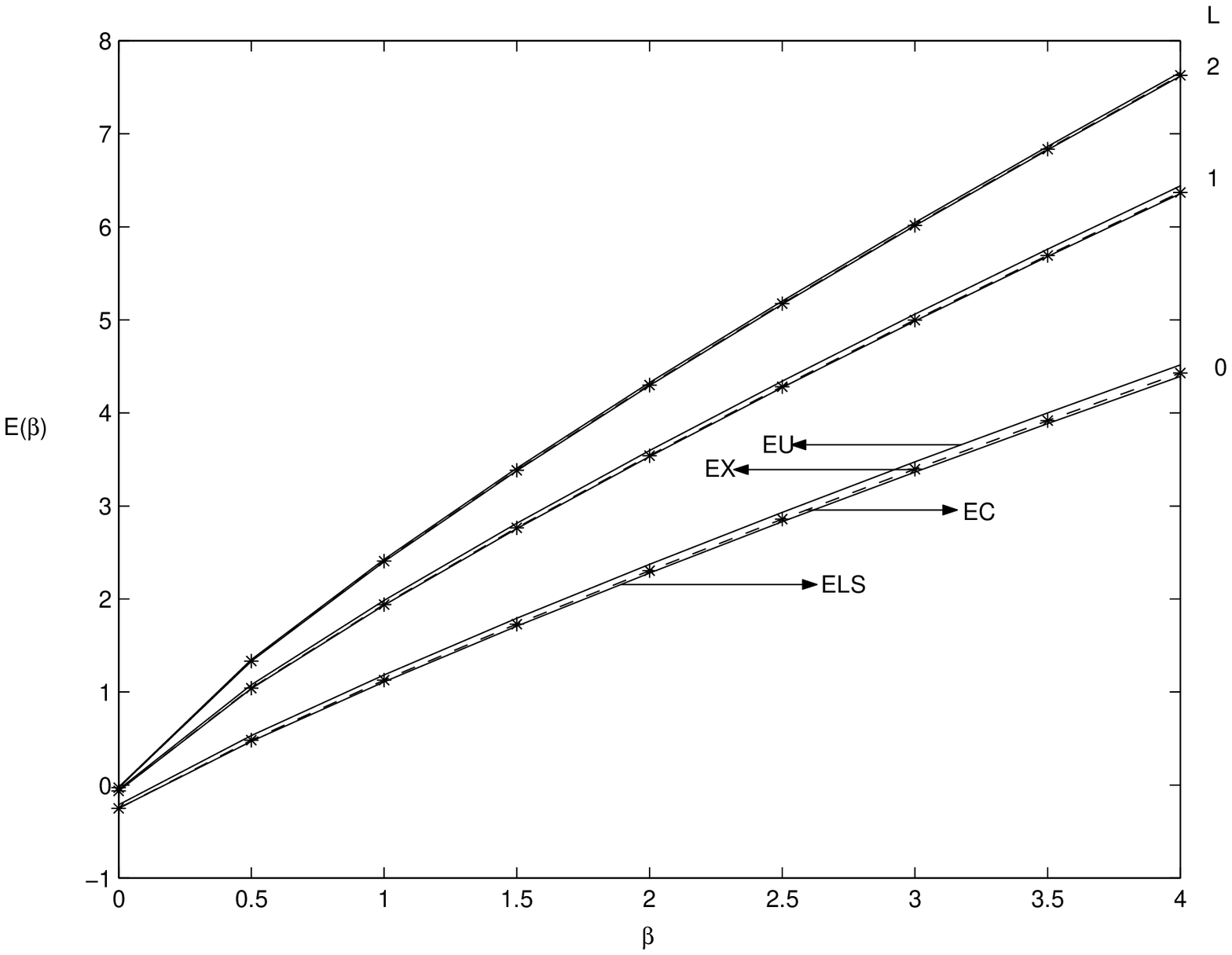,height=4in,width=5in,silent=}}}

\title{Figure 3.}

\nl The eigenvalues ${E}(\beta)$ of the Hamiltonian $H=-\Delta-1/r+\beta r^{0.5}$ for dimension $N = 3,$ $n=1,$ and $\ell=0,1,2$. The continuous curves show the upper bound EU given by the envelope formula (2.5) with $\nu = \mu = P_{1\ell}(0.5),$ and the lower bound ELS by the sum approximation 
given by the same formula but with $\nu = P_{1\ell}(0.5)$ and $\mu = P_{1\ell}(-1).$ The upper bound for $\ell=0$ is calculated using $\nu = P_{1\ell}^{U}(0.5)$ and $\mu = P_{1\ell}^{U}(-1)$ in formula (2.5).  The dashed curve $EC$ represents the upper bound by formula (3.6). The stars EX represent accurate numerical data.

\np
\references{1} 

\end